\documentclass[usenatbib,letterpaper,hyperref]{mn2e}
\usepackage[labelfont=bf,labelsep=period]{caption}
\usepackage[totalwidth=480pt, totalheight=680pt]{geometry}
\usepackage{hyperref}
\usepackage{times}
\usepackage{epsfig}
\usepackage{amsmath}
\usepackage{amssymb}
\usepackage{url}
\usepackage{aas_macros}
\DeclareRobustCommand{\VAN}[3]{#2}
\let\VANthebibliography\thebibliography
\def\thebibliography{\DeclareRobustCommand{\VAN}[3]{##3}\VANthebibliography}

\def\eqsim{\mathrel{\raise0.35ex\hbox{$\scriptstyle =$}\kern-0.6em
    \lower0.40ex\hbox{{$\scriptstyle \sim$}}}}
\def\gtrsim{\mathrel{\raise0.35ex\hbox{$\scriptstyle >$}\kern-0.6em
    \lower0.40ex\hbox{{$\scriptstyle \sim$}}}}
\def\lesssim{\mathrel{\raise0.35ex\hbox{$\scriptstyle <$}\kern-0.6em
    \lower0.40ex\hbox{{$\scriptstyle \sim$}}}}
\def\HI{H\,{\sc i}}
\def\AA{ALFALFA}
\def\HIWF{H\,\textsc{i}\,WF}
\def\HIMF{H\,\textsc{i}\,MF}
\def\EAGLECDM{EAGLE25\nobreakdash-CDM}
\def\EAGLEWDM{EAGLE25\nobreakdash-WDM\nobreakdash-$1.5\,\mathrm{keV}$}
\def\EMOSAICSCDM{EAGLE34\nobreakdash-CDM}
\def\EMOSAICSWDM{EAGLE34\nobreakdash-WDM\nobreakdash-$7\,\mathrm{keV}$}

\title[WDM and 21-cm surveys]{A warm dark matter cosmogony may yield more low-mass galaxy detections in \texorpdfstring{21\nobreakdash-cm}{21-cm} surveys than a cold dark matter one}
\author[K. A. Oman~et~al.]{Kyle A. Oman$^{1,2,3}$\thanks{kyle.a.oman@durham.ac.uk}, Carlos S. Frenk$^{1,3}$, Robert A. Crain$^{4}$, Mark R. Lovell$^{1,3,5}$ \newauthor and Joel Pfeffer$^{6}$\\
$^{1}$ Institute for Computational Cosmology, Durham University, South Road, Durham DH1 3LE, United Kingdom\\
$^{2}$ Centre for Extragalactic Astronomy, Durham University, South Road, Durham DH1 3LE, United Kingdom\\
$^{3}$ Department of Physics, Durham University, South Road, Durham DH1 3LE, United Kingdom\\
$^{4}$ Astrophysics Research Institute, Liverpool John Moores University, Liverpool, United Kingdom\\
$^{5}$ Centre for Astrophysics and Cosmology, Science Institute, University of Iceland, Dunhaga 5, 107 Reykjav\'{i}k, Iceland\\
$^{6}$ Centre for Astrophysics \& Supercomputing, Swinburne University, Hawthorn, VIC 3122, Australia
}
\date{\today}

\begin{document}
\label{firstpage}
\maketitle

\begin{abstract}
  The 21\nobreakdash-cm spectral line widths, $w_{50}$, of galaxies are an approximate tracer of their dynamical masses, such that the dark matter halo mass function is imprinted in the number density of galaxies as a function of $w_{50}$. Correcting observed number counts for survey incompleteness at the level of accuracy needed to place competitive constraints on warm dark matter (WDM) cosmological models is very challenging, but forward-modelling the results of cosmological hydrodynamical galaxy formation simulations into observational data space is more straightforward. We take this approach to make predictions for an \AA{}-like survey from simulations using the EAGLE galaxy formation model in both cold (CDM) and WDM cosmogonies. We find that for WDM cosmogonies more galaxies are detected at the low-$w_{50}$ end of the 21\nobreakdash-cm velocity width function than in the CDM cosmogony, contrary to what might na{\"i}vely be expected from the suppression of power on small scales in such models. This is because low-mass galaxies form later and retain more gas in WDM cosmogonies (with EAGLE). While some shortcomings in the treatment of cold gas in the EAGLE model preclude placing definitive constraints on WDM scenarios, our analysis illustrates that near-future simulations with more accurate modelling of cold gas will likely make strong constraints possible, especially in conjunction with new 21\nobreakdash-cm surveys such as WALLABY.
\end{abstract}

\section{Introduction}
\label{sec:intro}

The number density of dark haloes as a function of their mass, or of their (closely related) maximum circular velocity ($v_\mathrm{max}$), is a fundamental prediction of cosmological models with dark matter. The $\Lambda$CDM cosmogony predicts an approximately power-law subhalo $v_\mathrm{max}$ function with a slope of about $-3$ \citep[e.g.][]{2015MNRAS.454.1798K,2016MNRAS.457.3492H}. Testing this prediction is particularly interesting at the low-velocity end ($v_\mathrm{max}\lesssim 80\,\mathrm{km}\,\mathrm{s}^{-1}$) where warm dark matter \citep[WDM;][]{1983ApJ...274..443B,1986ApJ...304...15B,2001ApJ...556...93B} cosmogonies predict a flattening (followed by a cut-off at even lower $v_\mathrm{max}$) due to a truncation of the matter power spectrum leading to suppressed and delayed structure formation on small scales.

The subhalo $v_\mathrm{max}$ function is not directly observable, so we must turn to observable proxies. The most direct approach is to choose a measurement that can be interpreted as a dynamical mass tracer. One such measurement is the width of the 21\nobreakdash-cm emission line of neutral hydrogen (\HI{}). Under optimistic assumptions, the spectral line width (for example the full width at half-maximum, $w_{50}$) is related to $v_\mathrm{max}$ by a simple geometric correction: $w_{50}\approx 2v_\mathrm{max}\sin i$ where $i$ is the inclination angle. The first measurements of the \HI{} velocity width function \citep[hereafter \HIWF{},][see also \citealp{2011ApJ...742...16T,2015MNRAS.454.1798K}]{2009ApJ...700.1779Z,2010MNRAS.403.1969Z,2011ApJ...739...38P} found a flattening of the slope at the low-width end that could be interpreted as a consequence of WDM \citep[][but see \citealp{2015MNRAS.454.1798K}]{2009ApJ...700.1779Z,2011ApJ...739...38P,2017MNRAS.470.1542S}, but the current consensus is that the flattening is mostly due to the fact that the \HIWF{} is preferentially a tracer of the subhalo $v_\mathrm{max}$ function of gas-rich galaxies \citep{2013MNRAS.431.1366S,2013ApJ...766..137O,2017ApJ...850...97B,2019MNRAS.482.5606D}, rather than that of the global galaxy (or dark matter halo) population. However, as we will see below, provided that the bias towards gas-rich galaxies can be modeled in sufficient detail, the measured \HIWF{} remains sensitive to a cutoff in the matter power spectrum like that due to a thermal relic with a mass of about $10\,\mathrm{keV}$ or less.

The conventional approach \citep[e.g.][]{2009ApJ...700.1779Z,2010MNRAS.403.1969Z,2015A&A...574A.113P} to constrain the $v_\mathrm{max}$ function (often termed the `velocity function') observationally is to assign $v_\mathrm{max}$ values to surveyed galaxies based on their \HI{} line widths, such as by assuming $w_{50}=2v_\mathrm{max}\sin i$ or a more sophisticated but conceptually similar relationship. Once this is done the number density of surveyed galaxies is corrected for the limited sensitivity of the survey to estimate the intrinsic number density of galaxies as a function of $v_\mathrm{max}$. \citet{2017ApJ...850...97B} and \citet[][see also \citealp{2016MNRAS.455.3841B,2024ApJ...964..135S}]{2019MNRAS.482.5606D} have discussed some of the challenges of this approach, including: that the limited extent of \HI{} discs in low-mass galaxies often fails to reach the radii where the maximum of the circular velocity curve is reached; that many low-mass galaxies have \HI{} reservoirs too faint to be detected by current surveys; that \HI{} is increasingly pressure-supported (rather than rotation-supported) in lower-mass galaxies. We explore some further challenges in this work.

Rather than attempting to transform the \HIWF{} into an inferred measurement of a $v_\mathrm{max}$ function suitable for comparison with `raw' theoretical predictions, theoretical results can be recast into direct predictions for the \HIWF{} for comparison with `raw' observations, as advocated by \citet{2013ApJ...766..137O} and \citet{2022MNRAS.509.3268O}. One way to approach this is to predict line widths from $v_\mathrm{max}$ using essentially the reverse of the procedure described above: a variation on the theme of $w_{50}=2v_\mathrm{max}\sin i$ \citep[e.g.][]{2009ApJ...698.1467O,2011ApJ...739...38P,2015MNRAS.454.1798K,2019MNRAS.488.5898C,2023MNRAS.522.4043B}. This makes some of the challenges outlined above easier to deal with -- for example, the extent of the \HI{} disc can be folded into the construction of a spectrum from a circular velocity curve. However, such efforts often result in a population of model galaxies with gas kinematics and therefore spectra that are unrealistically regular and symmetric.

A promising alternative is to measure the \HI{} spectrum of galaxies in cosmological hydrodynamical simulations as this should more accurately capture the diverse morphological and kinematic irregularities that shape the spectra of real galaxies -- at least as far as these are accurately captured by the simulation in question. Some initial efforts along these lines have been made by \citet{2016MNRAS.463L..69M}, \citet{2017ApJ...850...97B} and \citet{2019MNRAS.482.5606D} and show that such a simulated survey can reproduce the observed \HIWF{} quite accurately -- and certainly more accurately than predicting line widths from $v_\mathrm{max}$ using an analytic prescription for the same samples of simulated galaxies.

In this work, we develop this methodology further by carrying out the first direct survey of a hydrodynamical simulation volume equivalent to that of the \AA{} 21\nobreakdash-cm survey (with the caveat that some tiling of periodic simulation volumes was required). We use a suite of CDM and WDM simulations to show that the effects of WDM can sometimes be counter-intuitive in this context, in the sense that they can lead to increased source counts for low-$w_{50}$ galaxies where we would na\"{i}vely expect fewer due to the suppression of the formation of low-mass haloes in WDM cosmogonies, and explain the reasons for this. We discuss the prospects of using the \HIWF{} as a constraint on WDM cosmogonies -- these are good, provided that a sufficiently faithful model of the galaxy population contributing to the measurement is realised in the simulations.

This paper is structured as follows. In Sec.~\ref{sec:data} we describe the simulation datasets and survey catalogues used in our analysis. In Sec.~\ref{sec:methods} we describe our mock survey methodology. Our mock-observed \HIWF{}s in CDM and WDM cosmogonies are presented and analysed in Sec.~\ref{sec:results}. We discuss advantages of mock surveys based on hydrodynamical simulations over methodologies requiring assumptions loosely of the form $w_{50}\approx v_\mathrm{max}\sin i$ in Sec.~\ref{sec:discussion}. Finally, we summarise our findings and likely future advances as new simulations and 21\nobreakdash-cm surveys become available in Sec.~\ref{sec:conclusions}.

\section{Data}
\label{sec:data}

\subsection{Simulations -- EAGLE and variations}
\label{subsec:data-sims}

We make use of four cosmological hydrodynamical galaxy formation simulations. The simulations were carried out with the EAGLE galaxy formation model, which uses the pressure-entropy formulation of smoothed-particle hydrodynamics \citep{2013MNRAS.428.2840H} with improved time-stepping criteria \citep{2012MNRAS.419..465D} and switches for artificial viscosity \citep{2010MNRAS.408..669C} and artificial conduction \citep{2008JCoPh.22710040P}. It implements radiative cooling \citep{2009MNRAS.393...99W}, star formation \citep{2004ApJ...609..667S,2008MNRAS.383.1210S}, stellar evolution and chemical enrichment \citep{2009MNRAS.399..574W}, black hole growth \citep{2005MNRAS.364.1105S,2015MNRAS.454.1038R}, energetic stellar \citep{2012MNRAS.426..140D} and black hole \citep{2009MNRAS.398...53B} feedback, and cosmic reionization \citep{2001cghr.confE..64H,2009MNRAS.399..574W}.

All four simulations were performed with mass resolution $8\times$ better than the flagship Ref-L100N1504 EAGLE simulation, yielding a baryon particle mass of $m_{\rm g} = 2.26\times 10^5\,{\rm M}_\odot$, and hence adopt the `Recal' model \citep[see][for further discussion]{2015MNRAS.446..521S}. The first, which we consider to be our fiducial point of reference, is a $(25\,\mathrm{Mpc})^3$ volume from the EAGLE suite run with a CDM cosmogony \citep[that advocated by the][]{2014A&A...571A..16P}, with identifier Recal-L0025N0752. We will refer to this as the `EAGLE25 cold dark matter' (or `\EAGLECDM{}') simulation. The second simulation that we use, introduced by \citet{2020MNRAS.498.1050B} and also examined in the recent study of \citet{2023arXiv231100041M}, is a $(25\,h^{-1}\,\mathrm{Mpc})^3\approx(34\,\mathrm{Mpc})^3$ volume followed at the same resolution, adopting the same cosmogony and the same Recal parameters for the EAGLE galaxy formation model. (This simulation also includes the E\nobreakdash-MOSAICS globular cluster formation and evolution model of \citealp{2018MNRAS.475.4309P,2019MNRAS.486.3134K}, but this is immaterial as it does not influence any of the galaxy properties.) We label this simulation `EAGLE34 cold dark matter' (or `\EMOSAICSCDM{}'). The matched resolutions, cosmogonies and galaxy formation models of these simulations offer a useful opportunity to control for cosmic variance effects using independent volumes.

The third and fourth simulations that we use are WDM counterparts to the first two. In each case exactly the same galaxy formation model as in the CDM simulation is used, and the initial conditions are also created to have the same phases \citep{2013MNRAS.434.2094J}, but with their power spectra modified to model the effect of WDM. The counterpart of the \EAGLECDM{} simulation uses the transfer function approximation of \citet{2001ApJ...556...93B} to represent the effect of a $1.5\,\mathrm{keV}$ thermal relic dark matter candidate. We choose this model as a relatively extreme case \citep[e.g. it is incompatible with structure formation constraints:][]{2021MNRAS.506.5848E,2021ApJ...917....7N} that is expected to emphasize any differences with respect to the CDM simulations. The model's half-mode wavenumber is $k_\mathrm{hm}=9.88\,\mathrm{Mpc}^{-1}$, and its half-mode mass is $M_\mathrm{hm}=5.3\times 10^{9}\,\mathrm{M}_\odot$ \citep{2016MNRAS.455..318B}. We label this `EAGLE25 $1.5\,\mathrm{keV}$ thermal relic' or simply `\EAGLEWDM{}'.

The counterpart of the \EMOSAICSCDM{} simulation has a power spectrum modified to mimic the effect of a $7.1\,\mathrm{keV}$ sterile neutrino dark matter candidate with a mixing angle $\sin(2\theta)=2\times10^{-11}$ (Meshveliani~et~al., in preparation). This is the `warmest' sterile neutrino model that is compatible with the X\nobreakdash-ray emission line observed in galaxies and clusters and tentatively associated with dark matter decay radiation \citep{2014PhRvL.113y1301B,2014ApJ...789...13B}. The sterile neutrino distribution function was computed using the implementations of \citet{2008JCAP...06..031L} and \citet{2016MNRAS.461...60L}, and the matter power spectrum was calculated with a modified version of the {\sc camb} Boltzmann-solver \citep{2000ApJ...538..473L}. We label this simulation `EAGLE34 $7\,\mathrm{keV}$ sterile neutrino $\sin(2\theta)=2\times10^{-11}$', or simply `\EMOSAICSWDM{}'.

Haloes are identified in all cases with a friends-of-friends algorithm with a linking length equal to $0.2$~times the mean inter-particle spacing \citep{1985ApJ...292..371D}, with gravitationally bound subhaloes subsequently identified using the \textsc{subfind} algorithm \citep{2001MNRAS.328..726S,2009MNRAS.399..497D}. In WDM cosmogonies, the truncation of the power spectrum causes artificial fragmentation of filaments in N-body simulations that could be spuriously identified as subhaloes. We remove these from halo catalogues following the approach of \citet{2014MNRAS.439..300L}. Such spurious subhaloes only outnumber `real' subhaloes at maximum circular velocities $v_\mathrm{max}\lesssim 15\,\mathrm{km}\,\mathrm{s}^{-1}$. In this regime subhaloes containing atomic hydrogen -- which are our main focus in this work -- are extremely rare.

\begin{figure}
  \includegraphics[width=\columnwidth]{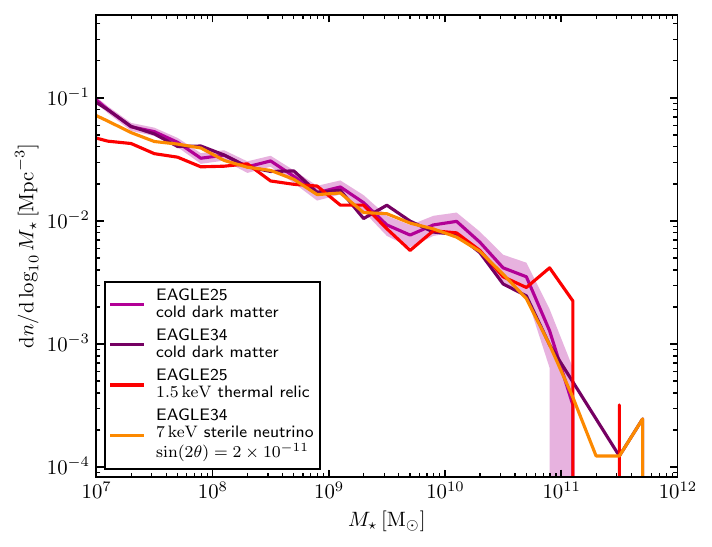}
  \caption{Galaxy stellar mass functions of simulations used in this work, for all galaxies with $M_\star>10^{7}\,\mathrm{M}_\odot$ (approximately $50$ stellar particles). Indicative counting uncertainties are illustrated with the shaded band for the \EAGLECDM{} simulation (these are slightly smaller in the larger EAGLE34 simulation volumes). The stellar mass functions agree closely, except at low stellar masses ($M_\star\lesssim 10^{8}\,\mathrm{M}_\odot$) where galaxy formation is suppressed in WDM cosmogonies.}
  \label{fig:gsmf}
\end{figure}

As a check that the `Recal' calibration of the EAGLE model remains applicable in the WDM cosmogonies that we consider, we show the galaxy stellar mass function in the $4$ simulations listed above in Fig.~\ref{fig:gsmf}. Outside of the regime where warm dark matter suppresses the formation of structure ($M_\star\lesssim 10^{8}\,\mathrm{M}_\odot$ for the relevant cosmogonies), the stellar mass functions agree very closely.

The EAGLE model does not directly compute the \HI{} fractions of simulation particles. Following \citet[][see also \citealp{2017MNRAS.464.4204C}]{2016MNRAS.456.1115B}, we partition neutral and ionized hydrogen following the prescription for self-shielding from the meta-galactic ionizing background radiation of \citet{2013MNRAS.430.2427R}, then subdivide the neutral hydrogen into atomic (\HI{}) and molecular (H$_2$) constituents using the empirically-calibrated pressure-dependent correction defined by \citet{2006ApJ...650..933B}.

\subsection{Observations -- The \AA{} survey}
\label{subsec:data-obs}

The \AA{} survey \citep{2005AJ....130.2598G} mapped about $7\,000\,\mathrm{deg}^2$ at 21\nobreakdash-cm wavelengths at declinations $0\leq\delta/\mathrm{deg}\leq36$ (avoiding the Galactic plane) using the Arecibo radio telescope. We use the $\alpha$\nobreakdash-$100$ \citep[full-survey;][see also \citealp{2011AJ....142..170H}]{2018ApJ...861...49H} release\footnote{\url{http://egg.astro.cornell.edu/alfalfa/data/}} of the \AA{} extragalactic source catalogue. We apply quality cuts as are conventional in studies of the \HIMF{} and \HIWF{} \citep[e.g.][]{2015A&A...574A.113P,2018MNRAS.477....2J,2022MNRAS.509.3268O}: we retain `Code 1' ($\mathrm{S}/\mathrm{N} > 6.5$) sources with \HI{} masses $10^{6}\leq M_\mathrm{HI}/\mathrm{M}_\odot\leq 10^{11}$, with line widths\footnote{Throughout this work we denote the natural logarithm $\log(\cdot)$, and the base-$10$ logarithm $\log_{10}(\cdot)$.} $1.2 \leq \log_{10}(w_{50}/\mathrm{km}\,\mathrm{s}^{-1}) < 2.95$, with recessional velocities $v_\mathrm{CMB}\leq 15\,000\,\mathrm{km}\,\mathrm{s}^{-1}$ and distances $D<200\,\mathrm{Mpc}$, that fall within the nominal survey footprint \citep[defined in tables~D1 \& D2 of][]{2018MNRAS.477....2J}. We use the updated distances of sources AGC~749235 and AGC~220210 reported by \citet{2022MNRAS.509.3268O}, and the TRGB distances for galaxies overlapping with the SHIELD sample reported by \citet{2021ApJ...918...23M}. These cuts result in a final sample of $21\,388$ galaxies.

\section{Methods}
\label{sec:methods}

\subsection{Mock \AA{}-like surveys}
\label{subsec:methods-mocks}

Whether a given galaxy is detected in the \AA{} survey depends on its \HI{} flux, $S_{21}$, and line width, $w_{50}$: a higher flux is of course easier to detect, and a narrower line width squeezes a given flux into a smaller number of channels in the detector, increasing the $\mathrm{S}/\mathrm{N}$. We therefore need to determine a flux and line width for each galaxy in a mock survey.

We begin by defining a mock survey volume with the same volume as our selection from the \AA{} survey (see Sec.~\ref{subsec:data-obs}). Since our simulations are of random cosmological volumes, we simplify the geometry of the survey to cover a circular aperture on the sky out to a maximum distance of $200\,\mathrm{Mpc}$. The \AA{} survey footprint \citep[][tables~D1 \& D2]{2018MNRAS.477....2J} subtends $6517.7\,\mathrm{deg}^{2}$, corresponding to an opening angle of $46.8\,\mathrm{deg}$ from the axis for our circular survey footprint. Since the survey volume is $339$ or $134$ times larger than the $(25\,\mathrm{Mpc})^{3}$ or $(34\,\mathrm{Mpc})^{3}$ simulations volumes, we replicate the periodic simulation volumes such that the survey volume is completely covered. We note that given the relative volumes and the opening angle, there is no advantage in terms of minimizing aliasing effects of choosing any particular survey axis orientation with respect to the simulation volume Cartesian axes. Given the potentially large number of copies of a given simulated galaxy that could appear in our mock surveys, we have carefully checked that aliasing effects are not unduly driving or influencing any of our conclusions. The main effect of tiling a limited volume is that the rarest (high-\HI{} mass, high $w_{50}$) galaxies that appear in the real survey are not sampled in the mocks. Since our focus is on the low-mass and low-velocity width regimes, this is of no concern. This focus also means that the galaxies that we are most interested in appear copied only up to about $10$ times in our mock catalogues, because they are faint and therefore undetected at distances beyond about $60\,\mathrm{Mpc}$. Finally, we note that no galaxies appear multiple times as exact copies since each instance of a galaxy has a different distance and is seen from a different viewing angle (and therefore has different $S_{21}$ and $w_{50}$, as explained below) than its aliases.

We next catalogue every subhalo in the mock survey volume that has $M_\mathrm{HI}>10^{6}\,\mathrm{M}_\odot$ and $M_{\star}>0$, and measure $S_{21}$ and $w_{50}$ for each. The flux is simply measured as:
\begin{equation}
  \frac{S_{21}}{\mathrm{Jy}\,\mathrm{km}\,\mathrm{s}^{-1}}=\left(2.36\times10^5\right)^{-1}\frac{M_\mathrm{HI}}{\mathrm{M_\odot}}\left(\frac{D}{\mathrm{Mpc}}\right)^{-2}, \label{eq:himass}
\end{equation}
where $M_\mathrm{HI}$ is the sum of the \HI{} masses of gas particles identified by \textsc{subfind} as bound to the subhalo and $D$ is the distance from the mock observer.

\begin{figure*}
  \includegraphics[width=\textwidth]{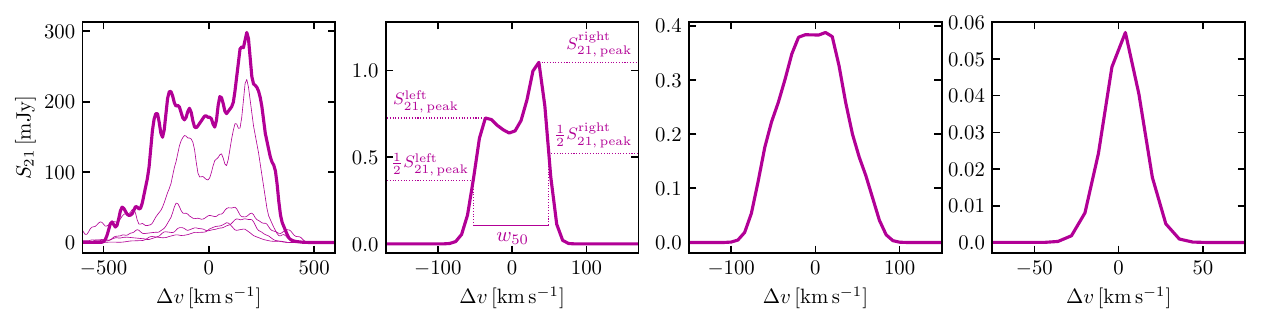}
  \caption{Sample 21\nobreakdash-cm spectra from the mock survey of the \EAGLECDM{} simulation. \emph{First panel:} a massive ($v_\mathrm{max}\approx 350\,\mathrm{km}\,\mathrm{s}^{-1}$) galaxy. The heavy line shows the spectrum of the nearest instance of the galaxy in the survey. Because the periodic simulation volume is replicated to cover the survey volume, the same galaxy appears more than once, but is seen at different distances and from different viewing angles each time -- spectra of other realisations of this galaxy are shown with thinner lines, illustrating that each instance has a different total flux and spectral line width. \emph{Second panel:} a spectrum of a lower-mass galaxy ($v_\mathrm{max}\approx 135\,\mathrm{km}\,\mathrm{s}^{-1}$) with a clear `double-horn' shape. The peak flux densities on the approaching and receding sides, $S_{21,\,\mathrm{peak}}^\mathrm{left}$ and $S_{21,\,\mathrm{peak}}^\mathrm{right}$, and the locations where half of their amplitudes are crossed, are marked with dotted lines. The full-width at half-maximum $w_{50}$ is marked with the thin solid line. \emph{Third panel:} a spectrum of a yet lower mass galaxy ($v_\mathrm{max}\approx 70\,\mathrm{km}\,\mathrm{s}^{-1}$) with an approximately flat-topped spectral shape. \emph{Fourth panel:} a spectrum of an even lower mass galaxy ($v_\mathrm{max}\approx 35\,\mathrm{km}\,\mathrm{s}^{-1}$) with an approximately Gaussian spectral shape.}
  \label{fig:spectra}
\end{figure*}

The calculation of the \HI{} line width is more involved. We first create a mock spectrum of each galaxy using the mock-observing tool \textsc{martini}\footnote{\url{https://github.com/kyleaoman/martini}, version~2.0.2.} \citep{2019MNRAS.482..821O,2019ascl.soft11005O,2024JOSS....9.6860O}. Each gas particle contributes a Gaussian spectral line profile to the spectrum with a centre determined by its velocity along the line of sight and a width:
\begin{equation}
  \sigma=\sqrt{\frac{k_\mathrm{B}T}{m_\mathrm{p}}},
\end{equation}
where $k_\mathrm{B}$ is Boltzmann's constant, $T$ is the particle temperature and $m_\mathrm{p}$ is the proton mass. The typical gas temperature is $T\approx 8\times10^{3}\,\mathrm{K}$, corresponding to $\sigma\approx 8\,\mathrm{km}\,\mathrm{s}^{-1}$. The initial spectral channels have a $4\,\mathrm{km}\,\mathrm{s}^{-1}$ width, and we use $512$ channels such that every spectrum is sampled in its entirety. As in \AA{}, we then Hanning smooth the spectra for a final spectral resolution\footnote{The channel spacing of the `raw' \AA{} spectra is about $5.5\,\mathrm{km}\,\mathrm{s}^{-1}$, or about $11\,\mathrm{km}\,\mathrm{s}^{-1}$ after Hanning smoothing. However, we find no qualitative differences in our analysis or conclusions even if we skip the smoothing step and proceed with a spectral resolution of $4\,\mathrm{km}\,\mathrm{s}^{-1}$. We therefore surmise that the small difference in spectral resolution can be safely ignored.} of $8\,\mathrm{km}\,\mathrm{s}^{-1}$. Since \AA{} is a low-spatial resolution drift-scan survey with nearly-uniform sensitivity across the survey area we do not take the morphology of the sources into account (i.e. we produce only a spectrum, not a data cube), nor do we model the primary beam. Since we process each subhalo separately, we implicitly neglect any source confusion. We also do not add any simulated noise to the measurements since the selection criteria include $\mathrm{S}/\mathrm{N}>6.5$ -- for our purposes, noise has a negligible influence on full-width at half maximum line widths in this regime. Fig.~\ref{fig:spectra} shows some example spectra of galaxies of different masses. The first panel illustrates that replicas of a given simulated galaxy generally have significantly different spectra, both in terms of total flux and shape.

With spectra of all galaxies in the survey volume in hand, we measure line widths as the full width at half maximum of each spectrum ($w_{50}$), loosely mimicking the approach used in the \AA{} survey. Where a spectrum has local maxima on either side of the systemic velocity ($\Delta v=0$ in Fig.~\ref{fig:spectra}), we identify the maximum on the left and on the right side separately. If these two maxima have comparable amplitudes (within a factor of $2$) we treat them separately; in any other case we use the single global maximum of the spectrum. We then linearly interpolate between channels to locate the velocity where 50~per~cent of the peak is crossed (using the separate peak flux density values, if applicable). This process is illustrated in the second panel of Fig.~\ref{fig:spectra}. In rare cases where the half-maximum flux level is crossed more than twice (keeping in mind that we allow the halo finder to separate the particles belonging to galaxies from their satellites or nearby companions), we use the crossings with the minimum and maximum $\Delta v$ to measure the width.

Our process to measure $w_{50}$ differs slightly from that used in constructing the \AA{} catalogue. In particular, the decision of whether to use one or two peaks is informed by visual inspection of individual spectra, and polynomial (usually linear) fitting of the edge regions of the spectra is used to locate the bounds defining $w_{50}$ rather than the direct interpolation between channels in our approach. Our decision to simplify the process somewhat (in particular to avoid the visual inspection step) is justified -- we find that even relatively drastic changes to the method, such as using no interpolation but simply taking the nearest channel, do not lead to qualitative changes in our analysis or conclusions.

With $S_{21}$ and $w_{50}$ determined for each galaxy in the mock survey volume, we are in a position to evaluate which galaxies would actually be detected by an \AA{}-like survey. We model \AA{}'s bivariate selection function with the empirically-determined form given in \citet[][eq.~(A5), based on the approach of \citealp{2011AJ....142..170H}]{2022MNRAS.509.3268O}. We linearly interpolate between the $25$, $50$ and $90$~per~cent completeness levels at fixed $S_{21}$, continuing the slope of the interpolation to extrapolate to $0$ and $100$~per~cent completeness levels. We decide whether a given simulated galaxy is detected by drawing a random number from a uniform distribution between $0$ and $1$ and comparing it to the survey completeness at its $S_{21}$ and $w_{50}$ -- if the random number is less than the completeness, the galaxy is accepted as a detection.

In addition to $S_{21}$ and $w_{50}$ we also catalogue the position within the survey cone (nominally a right ascension, a declination and a distance, although the former two carry little meaning since our simulations are of random cosmological volumes), and the maximum of the circular velocity curve $v_\mathrm{circ}(r)=\sqrt{GM(<r)/r}$ determined by \textsc{subfind}, where $M(<r)$ is the mass enclosed within a spherical aperture of radius $r$ and $G$ is Newton's constant.

\subsection{Matching galaxies across simulation volumes}
\label{subsec:methods-matching}

The pairs of simulations at fixed side length  (\EAGLECDM{} \& \EAGLEWDM{}, and \EMOSAICSCDM{} \& \EMOSAICSWDM{}) share the same random phases defining their initial conditions, so the same large-scale structures form in the two simulations of each pair. Identifying a galaxy from each as the `same' object is therefore well defined. In our analysis we will make some comparisons using such one-to-one matched galaxies. We identify the `best match' of each galaxy in a CDM simulation as the galaxy in the corresponding WDM simulation that contains the largest number of dark matter particles that belong to the CDM realisation of the galaxy\footnote{Particles are labelled with an identifier that is determined before particle positions are perturbed according to the power spectrum in the initial conditions, so the `same' particles in the CDM and WDM simulations share an identifier.}. We then repeat the process to identify the `best match' of each WDM galaxy in the corresponding CDM simulation. Candidate pairs are accepted as a match only if they are each others' mutual best matches (i.e. the match is `bijective'). This inevitably leaves a large fraction of galaxies -- especially satellites and the low-mass galaxies whose formation WDM physics suppresses -- unmatched, but this does not pose any problem for the comparisons that we will make below.

\section{Results}
\label{sec:results}

\subsection{The subhalo \texorpdfstring{$v_\mathrm{max}$}{vmax} function}

\begin{figure}
  \includegraphics[width=\columnwidth]{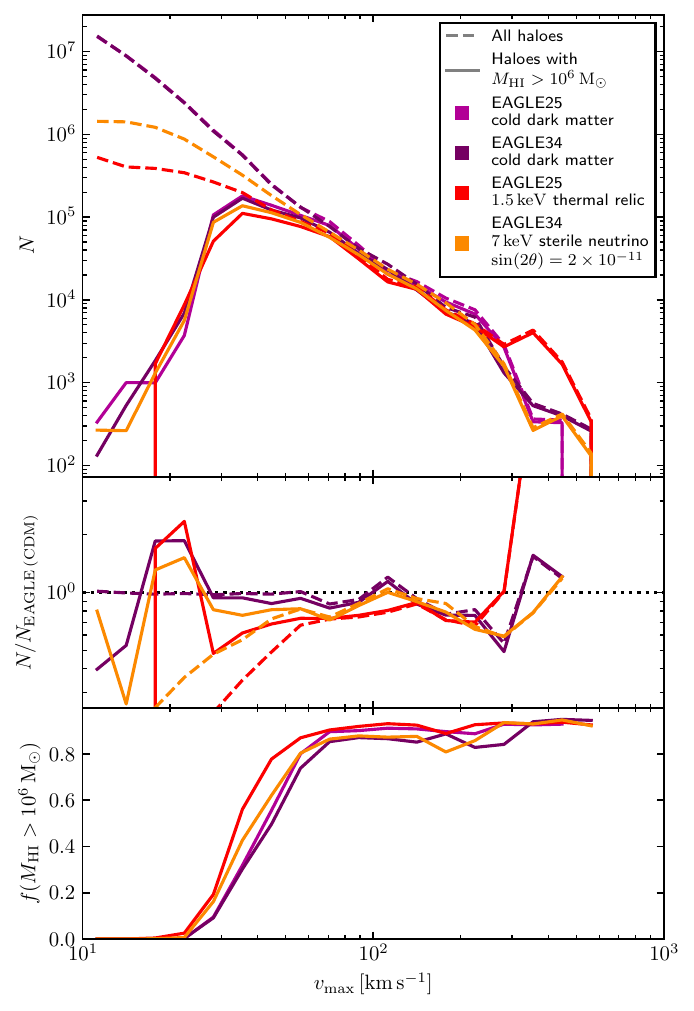}
  \caption{\emph{Upper panel:} the halo $v_\mathrm{max}$ function within a conical region with a volume equal to the \AA{} survey for the four simulations used in this work (\EAGLECDM{} -- light purple; \EMOSAICSCDM{} -- dark purple; \EAGLEWDM{} -- red; \EMOSAICSWDM{} -- orange). Spurious haloes in the WDM case have been removed (see Sec.~\ref{subsec:data-sims}). The bins in $v_\mathrm{max}$ have a width of $0.2\,\mathrm{dex}$. The dashed lines include all haloes within the conical volume, while the solid lines include only haloes with $M_\mathrm{HI}>10^{6}\,\mathrm{M}_\odot$. The suppression of galaxy formation on small scales in WDM cosmogonies is evident below $v_\mathrm{max}\approx 70\,\mathrm{km}\,\mathrm{s}^{-1}$. \emph{Middle panel:} ratio of the curves from the upper panel to the corresponding EAGLE (CDM) curves. \emph{Lower panel:} fraction of haloes that have $M_\mathrm{HI}>10^{6}\,\mathrm{M}_\odot$ as a function of $v_\mathrm{max}$. More low-mass ($v_\mathrm{max}\approx 40\,\mathrm{km}\,\mathrm{s}^{-1}$) haloes exceed this threshold in warmer dark matter cosmogonies.}
  \label{fig:vmax_function}
\end{figure}

The effect of WDM that we are aiming to constrain through observations is the suppression of the formation of low-mass galaxies. We therefore begin with a straightforward quantification of this suppression within our mock survey volumes -- the subhalo $v_\mathrm{max}$ function. In the upper panel of Fig.~\ref{fig:vmax_function} we show the number\footnote{Since we always work with the same fixed volume of $5.29\times10^{6}\,\mathrm{Mpc}^{3}$, we choose to show number counts rather than number densities.} of subhaloes (including `central' and `satellite' subhaloes) in the mock survey volume as a function of their maximum circular velocity for each of our four simulations. Reassuringly, the $v_\mathrm{max}$ functions of the two CDM simulations (light and dark purple dashed lines) agree very closely with differences of more than $10$~per~cent only occurring at $v_\mathrm{max}\gtrsim 100\,\mathrm{km}\,\mathrm{s}^{-1}$ (dashed dark purple line in centre panel) where sampling shot noise due to the limited volume of the simulations begins to become important\footnote{For instance, in the \EAGLECDM{} mock survey, a subhalo would be replicated about $339$ times, so at number counts less than $339\times100=3.4\times10^{4}$ we would expect shot noise at the level of $\approx (100)^{-1/2}=0.1$, i.e. $10$~per~cent. This is the abundance reached at about $100\,\mathrm{km}\,\mathrm{s}^{-1}$.}.

The two mock surveys of the WDM simulations have systematically fewer subhaloes at the low-$v_\mathrm{max}$ end. The `warmer' DM model (\EAGLEWDM{}) has a factor of about $40$ fewer haloes than the CDM simulations at $v_\mathrm{max}\approx 10\,\mathrm{km}\,\mathrm{s}^{-1}$, while the \EMOSAICSWDM{} simulation has a factor of $\approx 10$ fewer haloes than CDM at the same $v_\mathrm{max}$. The suppression of low-$v_\mathrm{max}$ subhaloes continues up to about $80$-$100\,\mathrm{km}\,\mathrm{s}^{-1}$ where differences are of about $10$-$20$~per~cent and begin to be difficult to distinguish from the shot noise in the subhalo $v_\mathrm{max}$ function due to the limited simulation volumes. \citet{2016MNRAS.455..318B} note that the suppression in the halo number density due to WDM is clear up to a scale of about $10M_\mathrm{hm}$ -- for the \EAGLEWDM{} simulation this is a subhalo mass scale of about $5\times10^{10}\,\mathrm{M}_\odot$, corresponding to a $v_\mathrm{max}$ scale of about $65\,\mathrm{km}\,\mathrm{s}^{-1}$, in good agreement with the curves in Fig.~\ref{fig:vmax_function}. We stress that at $v_\mathrm{max}\lesssim 80\,\mathrm{km}\,\mathrm{s}^{-1}$ the counting uncertainties (including accounting for tiling the simulations volumes) are much smaller than the amplitude of the suppression of number counts relative to CDM.

To take one step closer to the \AA{} observations that we wish to compare against, we also show the subhalo $v_\mathrm{max}$ function of galaxies with \HI{} masses of at least $M_\mathrm{HI}>10^6\,\mathrm{M}_\odot$ in the upper panel of Fig.~\ref{fig:vmax_function} (solid lines). At $v_\mathrm{max}$ of less than about $35\,\mathrm{km}\,\mathrm{s}^{-1}$ the number counts drop precipitously as most of these galaxies contain very little or no \HI{} gas. However, between $35\lesssim v_\mathrm{max}/\mathrm{km}\,\mathrm{s}^{-1}\lesssim 80$ the reduction in the number counts driven by WDM is still clearly evident (and the two CDM simulations still agree closely with each other). Interestingly, the fraction of galaxies with $M_\mathrm{HI}>10^{6}\,\mathrm{M}_\odot$ at fixed $v_\mathrm{max}$ in this range in $v_\mathrm{max}$ becomes higher with increasingly warm DM (lower panel of Fig.~\ref{fig:vmax_function}). This turns out to strongly affect the relative number counts of galaxies that are detected in our mock surveys; we will explore this in detail below.

\subsection{The \texorpdfstring{\HIMF{}}{HIMF}\ and \texorpdfstring{\HIWF{}}{HIWF}}
\label{subsec:himf_hiwf}

Since the detection of a galaxy in \AA{} depends primarily on its flux and line width, number counts as a bivariate function of $S_{21}$ and $w_{50}$ are, in a sense, the fundamental measurement obtained by the survey for population studies. Converting fluxes to masses (Eq.~(\ref{eq:himass})) helps to facilitate physical interpretation. Integrating the counts along the line-width axis yields the \HI{} mass function, while the orthogonal integral along the mass axis gives the \HI{} velocity width function. In a flux-limited survey number counts need to be corrected for the selection function of the survey in order to estimate number densities in a fixed volume \citep[e.g.][]{2003AJ....125.2842Z,2005MNRAS.359L..30Z}. Such estimates are subject to substantial statistical and systematic uncertainties \citep[e.g.][]{2018MNRAS.477....2J,2022MNRAS.509.3268O,2023MNRAS.522.4043B}. Going in the other direction -- drawing a flux-limited sample from a complete volume-limited catalogue -- is much more straightforward: it depends only on the assumed selection function in a simple way. Since volume-limited catalogues are trivial to construct from our mock survey volumes, we adopt an approach where we sample from these (Sec.~\ref{subsec:methods-mocks}) and make comparisons with the \AA{} survey in terms of observed number counts.

\begin{figure*}
  \includegraphics[width=\textwidth]{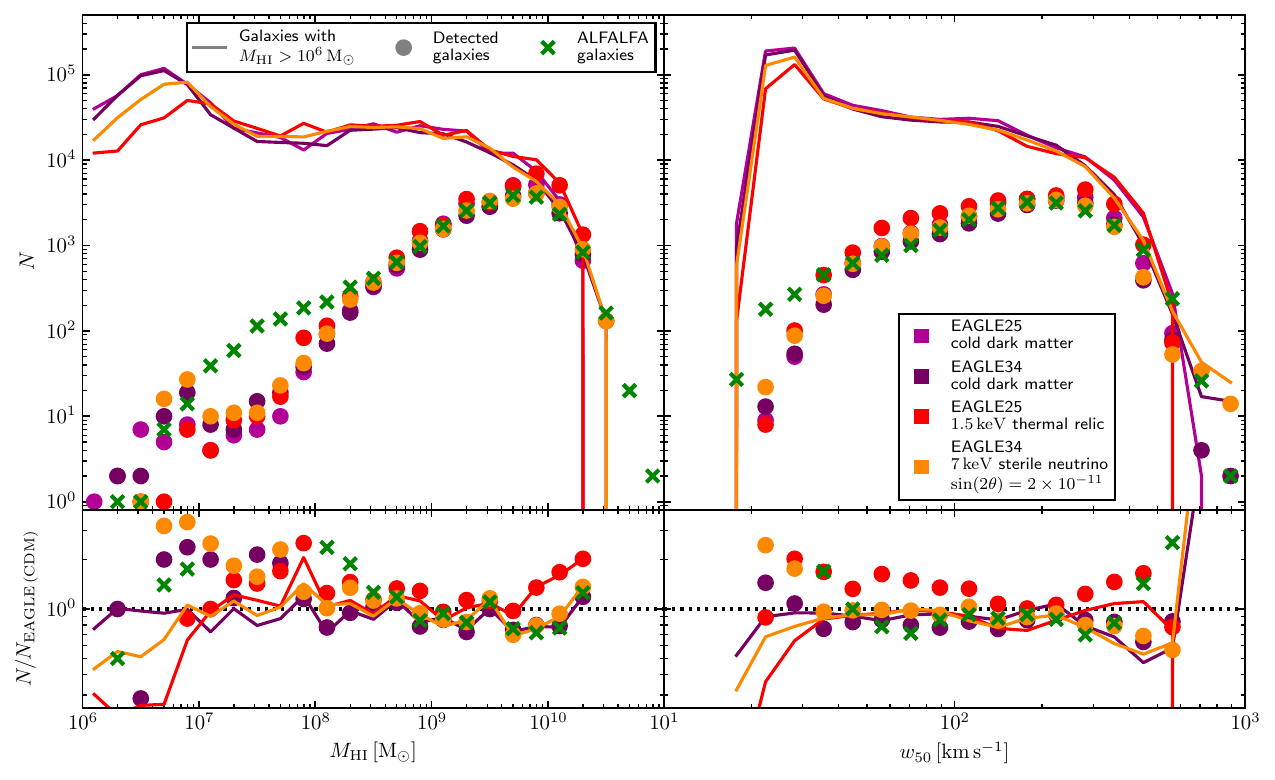}
  \caption{\emph{Upper left panel:} the \HIMF{} in the four simulations used in this work, and as observed by \AA{}. The solid curves show the counts of all galaxies in the mock survey cones, while points include only those that would be detected by an \AA{}-like survey. The colour coding (legend in upper right panel) is as in Fig.~\ref{fig:vmax_function}. The `bump' feature (solid curves) at low \HI{} mass ($M_\mathrm{HI}\lesssim 10^{7}\,\mathrm{M}_\odot$) has a numerical origin (see Sec.~\ref{subsec:himf_hiwf}). The green crosses show the number counts from the \AA{} extragalactic source catalogue. The bins in $M_\mathrm{HI}$ have a width of $0.1\,\mathrm{dex}$. \emph{Lower left panel:} ratio of the curves and points from the upper-left panel to the corresponding EAGLE (CDM) curve or points. \emph{Upper right panel:} as upper left panel, but for the \HIWF{}. The bins in $w_{50}$ have a width of $0.1\,\mathrm{dex}$. The suppression of number counts seen at low $v_\mathrm{max}$ in WDM cosmogonies is not evident at low $w_{50}$ -- in fact, the survey of the simulation with the `warmest' DM (EAGLE with $1.5\,\mathrm{keV}$ thermal relic) has a factor of $\approx 2$ more detections at $w_{50}\approx 70\,\mathrm{km}\,\mathrm{s}^{-1}$ than the surveys of CDM simulations. \emph{Lower right panel:} as lower right panel, but with ratios relative to EAGLE (CDM) corresponding to points and curves in the upper right panel.}
  \label{fig:hiwf}
\end{figure*}

In Fig.~\ref{fig:hiwf} we show the \HIMF{} (left panels) and \HIWF{} (right panels) of our four mock survey volumes. The solid lines show the number counts of all galaxies in the volumes with $M_\mathrm{HI}>10^{6}\,\mathrm{M}_\odot$ and $w_{50}>10\,\mathrm{km}\,\mathrm{s}^{-1}$, regardless of whether or not they are detected. Considering first the \HIMF{}, the two CDM simulations agree closely (with differences of about the amplitude expected from Poisson noise) across almost the entire range in $M_\mathrm{HI}$ shown. At the highest masses ($M_\mathrm{HI}>2\times 10^{10}\,\mathrm{M}_\odot$) the \EMOSAICSCDM{} curve turns up relative to the \EAGLECDM{} curve because the larger simulation volume of the former better samples the formation of these rare objects.

The two WDM simulations have very similar {\HIMF{}}s to the CDM simulations across most of the range in $M_\mathrm{HI}$, with no clear systematic trend with the `warmth' of the DM in the range $2\times 10^{7}\lesssim M_\mathrm{HI}/\mathrm{M}_\odot\lesssim 5\times 10^{9}$. In the most massive galaxies, the \EAGLEWDM{} model seems to cool substantially more gas than the other models and forms galaxies with very high central stellar densities (a point to which we will return below). At the low \HI{} mass end, number counts at fixed $M_\mathrm{HI}$ decrease for increasingly warm DM, reflecting the lower abundance of dark matter haloes in WDM cosmogonies -- the median $v_\mathrm{max}$ of a halo hosting a galaxy with $M_\mathrm{HI}=10^{7}\,\mathrm{M}_\odot$ is about $35\,\mathrm{km}\,\mathrm{s}^{-1}$ (in all $4$ simulations). While this is a potentially interesting difference between the predictions of CDM and WDM models, it occurs at an \HI{} mass scale below that where resolution-dependent numerical effects begin to occur in the EAGLE galaxy formation model \citep[$M_\mathrm{HI}\approx 5\times10^{7}\,\mathrm{M}_\odot$, see][sec.~4.2.1 for a detailed discussion -- we note that our selection excludes the `dark' haloes discussed in that work]{2017MNRAS.464.4204C}. Observed number counts in the \AA{} survey (green crosses; see Sec.~\ref{subsec:data-obs}) are also very low in this regime, with fewer than $30$ galaxies detected with $M_\mathrm{HI}<10^{7}\,\mathrm{M}_\odot$; this will be mitigated in the future by wider surveys of comparable sensitivity. Encouragingly, the observed number counts actually agree rather well with those in all simulations (coloured points) for \HI{} masses $\gtrsim 3 \times 10^{8}\,\mathrm{M}_\odot$ -- that is, the regime where resolution-dependent effects are not known to occur -- with the exception of the most massive galaxies, which are not reproduced by our relatively small simulation volumes.

We turn next to the \HIWF{} (right panels of Fig.~\ref{fig:hiwf}) which, through its closer connection to the dynamical masses of galaxies, is perhaps a more promising means to discriminate reliably between CDM and WDM cosmogonies. The most prominent feature in the number count distribution of all galaxies (solid curves) is a peak at $w_{50}$ of just over $20\,\mathrm{km}\,\mathrm{s}^{-1}$. The sharp decline towards lower line widths is due to the minimum line width imposed by thermal broadening of the \HI{} spectra -- almost all gas identified as \HI{} in the simulations has a temperature of about $8\times10^{3}\,\mathrm{K}$, corresponding to $w_{50}$ of about $2\sqrt{2\log(2)} (8\,\mathrm{km}\,\mathrm{s}^{-1})\approx 19\,\mathrm{km}\,\mathrm{s}^{-1}$. The sharp increase in the number counts from $w_{50}$ of about $40$ towards $20\,\mathrm{km}\,\mathrm{s}^{-1}$ is related to the bump in the \HIMF{} at $M_\mathrm{HI}\approx 5\times 10^{6}$-$10^{7}\,\mathrm{M}_\odot$ -- the median $M_\mathrm{HI}$ at a $w_{50}$ of $20\,\mathrm{km}\,\mathrm{s}^{-1}$ is $3$-$6\times 10^{6}\,\mathrm{M}_\odot$ (depending on the particular simulation), with a $16^\mathrm{th}$-$84^\mathrm{th}$~percentile scatter of about $0.6\,\mathrm{dex}$. In this extreme low-$w_{50}$ regime the suppression of number counts due to WDM physics is manifest (lower-right panel of Fig.~\ref{fig:hiwf}), but surprisingly it only appears at $w_{50}\lesssim 35\,\mathrm{km}\,\mathrm{s}^{-1}$. In Fig.~\ref{fig:vmax_function} we showed that number counts are suppressed in our WDM simulations at $v_\mathrm{max}\lesssim 80\,\mathrm{km}\,\mathrm{s}^{-1}$, and haloes with a given $v_\mathrm{max}$ are expected to contribute to the \HIWF{} at $w_{50}\lesssim 2v_\mathrm{max}$, so we might expect to see differences even up to $w_{50}\approx 160\,\mathrm{km}\,\mathrm{s}^{-1}$.

The result is even more surprising when we consider only those galaxies that are detected in our mock surveys (coloured points in right panels of Fig.~\ref{fig:hiwf}). In this case differences between CDM and WDM (mainly the more extreme \EMOSAICSWDM{} model) do persist to higher $w_{50}\gtrsim 100\,\mathrm{km}\,\mathrm{s}^{-1}$, but have the opposite sign to that which would na\"{i}vely be predicted -- more galaxies are detected in the mock survey of the WDM simulation than of the CDM simulations! We will detail the reasons for this in the next subsection, but first we briefly comment on the comparison of the {\HIWF{}}s from our mock surveys with that observed by \AA{}. These agree rather closely -- within about 20~per~cent or better in the case of the \EAGLECDM{}, \EMOSAICSCDM{} and \EMOSAICSWDM{} simulations -- across most of the range in $w_{50}$. At $w_{50}\lesssim 35\,\mathrm{km}\,\mathrm{s}^{-1}$ resolution-dependent effects in the simulations begin to become important, as mentioned above, so the departure here likely reflects a failure of the simulations to adequately resolve \HI{} gas physics in the lowest-mass galaxies\footnote{The agreement at $w_{50}<35\,\mathrm{km}\,\mathrm{s}^{-1}$, and also at $M_{\mathrm{HI}}<10^{8}\,\mathrm{M}_\odot$, is somewhat improved if simulated galaxies with $M_\star=0$ are included in the selection, but we note that all \AA{} sources included in the sample have optical counterparts. See \citet[][sec.~4.2.1]{2017MNRAS.464.4204C} for further discussion of the (largely unphysical) origin of these `dark' galaxies in the simulations.}. At the high-$w_{50}$ end the culprit is once again most plausibly the relatively small volume of our simulations precluding the formation of the rarest massive galaxies. This reasonably good agreement between the observed and mock-observed {\HIWF{}}s is far from trivial -- see for instance \citet[][fig.~4]{2023MNRAS.522.4043B} for an example of a galaxy formation model that fails rather spectacularly in its prediction for the \HIWF{}. We will discuss the origin of this agreement further in Sec.~\ref{subsec:simplemock}.

\subsection{Why more galaxies are detected in WDM models}

\begin{figure}
  \includegraphics[width=\columnwidth]{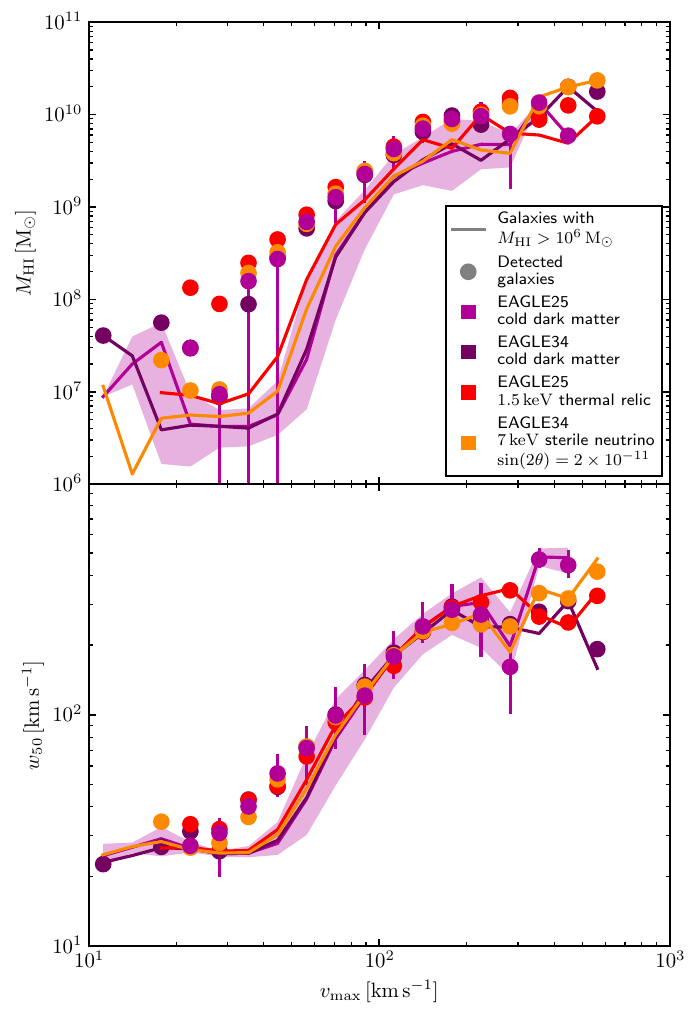}
  \caption{\emph{Upper panel:} median \HI{} mass of all galaxies with $M_\mathrm{HI}>10^{6}\,\mathrm{M}_\odot$ within the mock survey cones (solid lines) and of only those galaxies detected in the mock surveys (points), as a function of maximum circular velocity, $v_\mathrm{max}$. The upturn in the curves at the low-$v_\mathrm{max}$ end is an edge effect due to the cut in \HI{} mass. Colour coding (see legend) is as in Fig.~\ref{fig:vmax_function}. The interquartile scatter of the distributions is shown for the EAGLE (CDM) simulation as the shaded band and error bars; the scatter in the other simulations is very similar. \emph{Lower panel:} as the upper panel, but showing the median \HI{} line width $w_{50}$ as a function of $v_\mathrm{max}$.}
  \label{fig:medians}
\end{figure}

There is already a hint of the reason why more galaxies with line widths of about $40$-$100\,\mathrm{km}\,\mathrm{s}^{-1}$ are detected in our mock surveys of simulations with `warmer' DM in the lower panel of Fig.~\ref{fig:vmax_function}: at fixed $v_\mathrm{max}$ between about $20$-$70\,\mathrm{km}\,\mathrm{s}^{-1}$ the fraction of haloes with \HI{} masses greater than $10^{6}\,\mathrm{M}_\odot$ systematically increases for increasingly warm DM. We explore this further in Fig.~\ref{fig:medians}. In the upper panel, we plot the median \HI{} mass of galaxies as a function $v_\mathrm{max}$. The solid lines show the trends when all galaxies are included. Below $v_\mathrm{max}$ of about $100\,\mathrm{km}\mathrm{s}^{-1}$, the median \HI{} mass at fixed $v_\mathrm{max}$ systematically increases for increasingly warm DM. Considering only those galaxies that are detected in the mock surveys, we find that these are biased to higher median \HI{} masses than the global galaxy population. This is intuitive -- the surveys are biased towards the brightest sources.

The lower panel of Fig.~\ref{fig:medians} is similar to the upper panel, but shows the median $w_{50}$ at fixed $v_\mathrm{max}$. There is a trend for galaxies with $40<v_\mathrm{max}/\mathrm{km}\,\mathrm{s}^{-1}<80$ to have slightly higher median $w_{50}$ with increasingly warm DM; this is because their higher \HI{} masses (as shown in the upper panel) cause their gas discs to be somewhat more extended and therefore trace the rotation curves to slightly larger radii. As should be expected from the very tight correlation between \HI{} mass and size \citep[see][for a detailed discussion of its origin]{2019MNRAS.490...96S}, gas discs have the same sizes at fixed \HI{} mass in all of the simulations (within 15~per~cent or less), so we focus our attention on systematic offsets in \HI{} mass rather than size. Considering only those galaxies detected in our mock surveys, we find that at the low-$v_\mathrm{max}$ end (about $30<v_\mathrm{max}/\mathrm{km}\,\mathrm{s}^{-1}<70$) these are biased to wider line widths than the overall galaxy population. Unlike the detection bias in \HI{} mass, this runs contrary to intuition: all else being equal, a wider line has a lower signal-to-noise ratio than a narrow line and should therefore be more difficult to detect. This shows that the bias towards detecting galaxies with higher $M_\mathrm{HI}$ dominates over that towards detecting galaxies with lower $w_{50}$ at any given $v_\mathrm{max}$ in our mock surveys.

The offset between the curves in the upper panel of Fig.~\ref{fig:medians} could be interpreted either as a systematic difference in $M_\mathrm{HI}$ or in $v_\mathrm{max}$. That is, we could imagine either that a galaxy in a WDM simulation ends up with more \HI{} gas than its counterpart in the corresponding CDM simulation, or instead that it ends up with the same amount of \HI{} gas but a lower $v_\mathrm{max}$ (or a combination of both effects). Around $v_\mathrm{max}$ of $30\,\mathrm{km}\,\mathrm{s}^{-1}$, where the curves are nearly horizontal, it is clear that the driving effect must be a difference in $M_\mathrm{HI}$, but around $v_\mathrm{max}$ of $70\,\mathrm{km}\,\mathrm{s}^{-1}$ where the curves are steeper the situation is ambiguous. We resolve this ambiguity by using our individually matched pairs of haloes in WDM and corresponding CDM simulations (see Sec.~\ref{subsec:methods-matching} for details of our matching process).

\begin{figure}
  \includegraphics[width=\columnwidth]{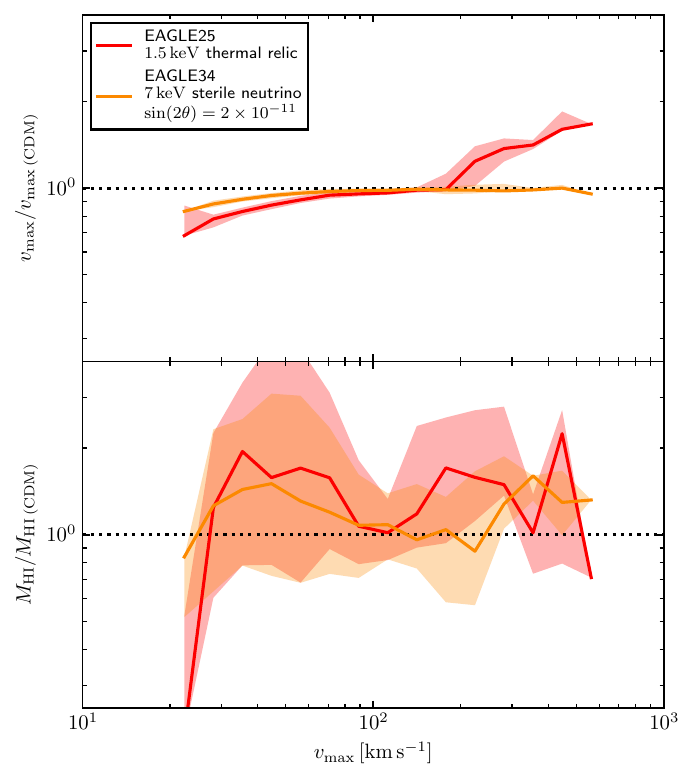}
  \caption{\emph{Upper panel:} the median maximum circular velocity $v_\mathrm{max}$ (solid line) and interquartile scatter (shaded band) of WDM haloes relative to the $v_\mathrm{max}$ of the matching halo from the corresponding CDM simulation. See Sec.~\ref{subsec:methods-matching} for details of the matching procedure. The $v_\mathrm{max}$ of low-$v_\mathrm{max}$ haloes is reduced by $10$-$30$~per~cent with little scatter in the WDM cosmogonies considered. The upturn in the \EMOSAICSWDM{} curve at $v_\mathrm{max}\gtrsim 200\,\mathrm{km}\,\mathrm{s}^{1}$ is due to high central baryon densities causing a central peak in the circular velocity curves of massive galaxies in the WDM simulation. \emph{Lower panel:} as upper panel, but comparing the \HI{} masses of WDM galaxies with those of their CDM counterparts. Low-mass galaxies have systematically higher gas masses by up to a factor of $2$, with considerable scatter. This is similar to the trend shown in the upper panel of Fig.~\ref{fig:medians}, but for matched haloes such that the systematic shift in $v_\mathrm{max}$ (upper panel of this figure) is removed.}
  \label{fig:matched}
\end{figure}

In Fig.~\ref{fig:matched} we show the median ratio of $v_\mathrm{max}$ of haloes in the two WDM simulations and the $v_\mathrm{max}$ of the matched CDM halo in the upper panels, and similarly the median ratio of $M_\mathrm{HI}$ in the lower panel. This reveals that both quantities are systematically different in the two simulations. $v_\mathrm{max}$ is systematically slightly lower in WDM (by $10$-$30$~per~cent) at the low-$v_\mathrm{max}$ end, with little scatter. This is a familiar effect due to the lower concentrations of WDM haloes at fixed halo mass \citep[see for example][fig.~4]{2023MNRAS.520.1567L}. $M_\mathrm{HI}$, on the other hand, is systematically higher by up to a factor of $2$, with large scatter. Most of the offset between the curves in the upper panel of Fig.~\ref{fig:medians} therefore comes from systematic differences in \HI{} mass, although systematic differences in $v_\mathrm{max}$ do play a small role.

The origin of the enhanced \HI{} mass of low-mass galaxies in the WDM simulations seems to be related to their assembly histories. Galaxies with $30<v_\mathrm{max}/\mathrm{km}\,\mathrm{s}^{-1}<50$, where the effect is most pronounced, form their stellar mass very late -- after $z=4$ (in \EMOSAICSWDM{}) or $z=3$ (in \EAGLEWDM{}), as opposed to from about $z=5.5$ in the CDM simulations. Yet, galaxies of a given $v_\mathrm{max}$ assemble almost exactly the same stellar mass in the CDM and corresponding WDM models. This lag in formation time persists up to about $v_\mathrm{max}=100\,\mathrm{km}\,\mathrm{s}^{-1}$, becoming smaller with increasing $v_\mathrm{max}$; this is also the value of $v_\mathrm{max}$ where the CDM and corresponding WDM simulations have about the same \HI{} mass per galaxy at fixed $v_\mathrm{max}$ (the difference in \HI{} mass grows again at larger $v_\mathrm{max}$). The delayed formation of structure at the low-mass end in WDM models, and the related delayed star formation in low-mass galaxies, likely allows galaxies to retain more gas over time. That galaxies near the cutoff scale in WDM form by direct collapse and accretion rather than mergers of smaller objects, which in CDM would likely have lost all of their gas through supernova feedback\footnote{Similar effects are present in the ETHOS model of self-interacting dark matter (SIDM), which exhibits a matter power spectrum cutoff qualitatively similar to WDM. An analysis of SIDM-CDM matched pairs at $z>6$ revealed systematically higher gas masses for galaxies in the SIDM model \citep{2019MNRAS.485.5474L}.}. A more definitive explanation would require a full merger tree analysis of the simulations which we defer to future work.

Although we are focused primarily on galaxies with $v_\mathrm{max}\lesssim 100\,\mathrm{km}\,\mathrm{s}^{-1}$ in this work, we feel compelled to comment on the conspicuous upturn of the \EAGLEWDM{} curve at $v_\mathrm{max}\gtrsim 200\,\mathrm{km}\,\mathrm{s}^{-1}$ in the upper panel of Fig.~\ref{fig:matched}. This comes from dense accumulations of stars and gas in the centres of massive galaxies in this simulation, causing a central peak in the circular velocity curve that much exceeds the circular velocity reached at larger radii. We have checked explicitly that no equivalent feature exists in a plot of $M_{200}/M_{200\,\mathrm{(CDM)}}$ versus $M_{200}$; $v_\mathrm{max}$ is simply no longer a good tracer of total dynamical mass in these galaxies. Since these objects are not our focus in this work, we will not dwell further on this detail.

\section{Discussion}
\label{sec:discussion}

In this section we compare our mock survey approach with a simplified approach to highlight some of its strengths and comment on its potential to produce constraints on WDM cosmological models.

\subsection{A simplified mock survey}
\label{subsec:simplemock}

As mentioned in Sec.~\ref{subsec:himf_hiwf}, that a galaxy formation model will reproduce the \HIWF{} even very approximately is far from a foregone conclusion. We explore some of the reasons for this in this section with the help of a simplified mock survey, using the \EAGLECDM{} simulation as an illustrative example.

We keep exactly the same setup described in Sec.~\ref{subsec:methods-mocks} except for the calculation of the line widths, $w_{50}$. We replace the entire calculation with a simple -- if often used -- approximation:
\begin{align}
  w_{50} = \sqrt{\left(2v_\mathrm{max}\sin i\right)^2 + \left(10\,\mathrm{km}\,\mathrm{s}^{-1}\right)^2}.\label{eq:vsini}
\end{align}
This assumes an inclined (by $i$) rotating disc that reaches an amplitude $v_\mathrm{max}$. The second term in the sum in quadrature reflects the fact that \HI{} gas is not perfectly dynamically cold but has an intrinsic velocity dispersion of about $10\,\mathrm{km}\,\mathrm{s}^{-1}$, which will dominate the line width once the rotation speed (or the inclination) becomes low enough.

\begin{figure*}
  \includegraphics[width=\textwidth]{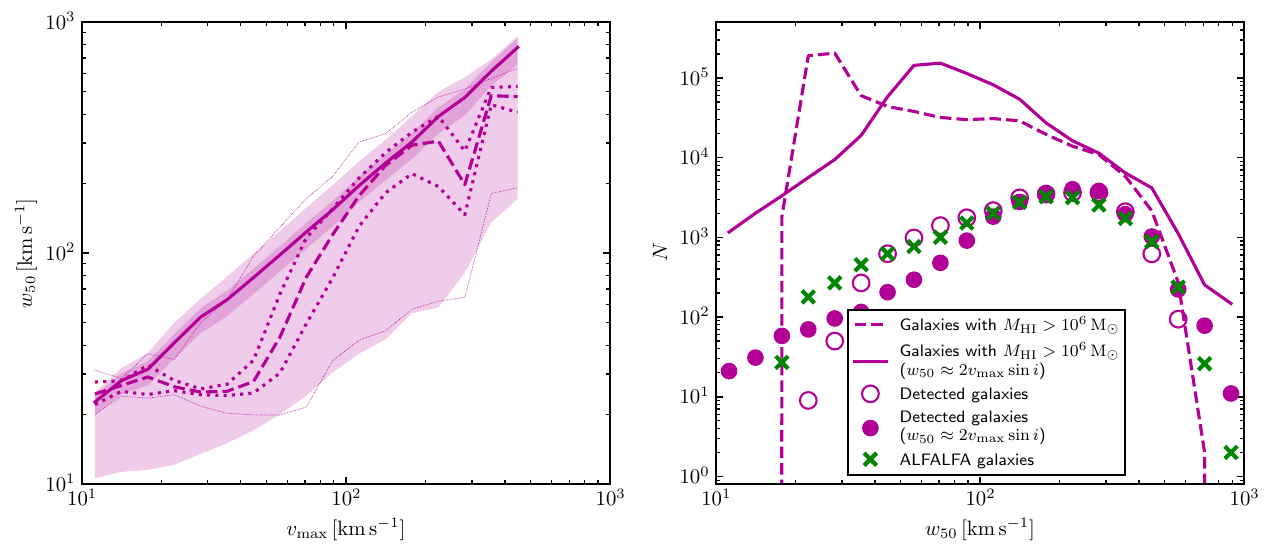}
  \caption{Comparison of a mock survey using our fiducial `observed' \HI{} line widths, $w_{50}$, with approximate line widths estimated as $w_{50}\approx 2v_\mathrm{max}\sin i$ (see Sec.~\ref{subsec:simplemock} for details), for the \EAGLECDM{} simulation. \emph{Left panel:} median $w_{50}$ as a function of $v_\mathrm{max}$ using `observed' (dashed line, repeated from Fig.~\ref{fig:medians}) and approximate (solid line) values of $w_{50}$. The interquartile scatter is shown with the heavier dotted lines and darker shaded region, respectively, and the thinner dotted lines and lighter shaded region show the $0.5^\mathrm{th}$-$99.5^\mathrm{th}$ percentile range. \emph{Right panel:} \HIWF{} using `observed' and approximate values of $w_{50}$. Lines are as in the left panel. The open symbols show the number of galaxies detected in the mock survey when the `observed' values of $w_{50}$ are used (repeated from the right panel of Fig.~\ref{fig:hiwf}) while the filled symbols show the same when approximate $w_{50}$ values are used. Number counts from the \AA{} catalogue are also shown for comparison. There are several, often competing effects (detailed in Sec.~\ref{subsec:simplemock}) that shape the trends in the right panel.}
  \label{fig:vsini}
\end{figure*}

We plot the median $w_{50}$ obtained in this way as a function of $v_\mathrm{max}$ in the left panel of Fig.~\ref{fig:vsini}, which displays the expected straight line with a slight upward inflection at the low-$v_\mathrm{max}$ end where the second term begins to contribute. For comparison we also show the curve using the mock-observed $w_{50}$ values (reproduced from Fig.~\ref{fig:medians}), which has a more complex shape.

We show the \HIWF{} for each of these cases in the right panel of Fig.~\ref{fig:vsini}. There are several, often competing effects driving the many differences in the figure:
\begin{enumerate}
\item In low-mass galaxies the gas disc does not sample the rotation curve all the way to $v_\mathrm{max}$, but only the central part where it is still rising. This causes galaxies that had $w_{50}$ of less than about $60\,\mathrm{km}\,\mathrm{s}^{-1}$ in the fiducial mock survey to have larger $w_{50}$ in the simplified mock survey -- this is the origin of the peak in the solid line in the right panel of Fig.~\ref{fig:vsini} at that line width.
\item The overestimated line widths in the simplified mock survey from (i) mean that more intrinsically faint galaxies have higher line widths, suppressing the fraction of galaxies that are detected at some line widths. For instance, even though there are a factor of several more galaxies with $w_{50}$ of about $60\,\mathrm{km}\,\mathrm{s}^{-1}$ in the simplified mock survey (solid line compared to dashed line), a factor of a few less are detected (solid points compared to open points) at the same line width.
\item In the simplified survey there is a long low-$w_{50}$ tail in the distribution at fixed $v_\mathrm{max}$ due to face-on galaxies, extending all the way to the velocity dispersion floor that we impose at $10\,\mathrm{km}\,\mathrm{s}^{-1}$. Galaxies in this tail have unrealistically narrow \HI{} lines (because the discs are assumed to be unrealistically symmetric and without warps or other irregular features common in observed galaxies), enabling them to satisfy our detection criteria at unrealistically large distances. This is part of the reason why between $w_{50}$ of $20$ and $70\,\mathrm{km}\,\mathrm{s}^{-1}$, for instance, the relative galaxy counts in the fiducial and simplified mock surveys change by almost two orders of magnitude (compare solid and dashed lines in right panel of Fig.~\ref{fig:vsini}), but the counts of detected galaxies varies by only about a factor of $5$ over the same interval in $w_{50}$. Even though the tail of the $w_{50}$ distribution has a low amplitude (e.g. the $0.1$ percentile of the distribution shown in the left panel of Fig.~\ref{fig:vsini}), the fraction of galaxies detected at a given $w_{50}$ is less than $1$~per~cent over much of the range in $w_{50}$, such that even a tiny fraction of affected galaxies can be important.
\end{enumerate}

\subsection{Implications for use of the \texorpdfstring{\HIWF{}}{HIWF} to constrain dark matter}

The above considerations make modelling the $w_{50}$ function to infer a $v_\mathrm{max}$ function an extremely challenging prospect. The first point was already identified by \citet[][see also \citealp{2019MNRAS.482.5606D}]{2017ApJ...850...97B}, but the other two have not, to our knowledge, been previously discussed in the literature. The mock survey based on a semi-analytic model of \citet{2023MNRAS.522.4043B} is similar to our simplified mock survey in some respects but accounts for the \HI{} sizes of galaxies and how these sample their rotation curves. The \HIWF{} that it predicts is more similar to that of our simplified mock survey than to our fiducial mock survey (although the agreement with \AA{} observations is worse than either!), suggesting that rotation curves that do not reach $v_\mathrm{max}$ are not the only effect at play.

Fundamentally, the issue is that given an observation of the \HI{} line width of a source, there is ambiguity in what value of $v_\mathrm{max}$ should be assigned to it. A common assumption \citep[e.g.][]{2010MNRAS.403.1969Z,2015A&A...574A.113P} is precisely that of our simplified mock survey (Eq.~(\ref{eq:vsini})). The left panel of Fig.~\ref{fig:vsini} illustrates that a realistic mapping $v_\mathrm{max}(w_{50})$ is likely not single-valued \citep[see also][fig.~3, and \citealp{2019MNRAS.482.5606D}, fig.~4]{2017ApJ...850...97B}. Fig.~\ref{fig:vsini} as a whole shows that (erroneously) assuming a relation similar to Eq.~(\ref{eq:vsini}) is likely to have severe consequences for constraints on the galaxy population and, by extension, on the halo $v_\mathrm{max}$ function.

The forward-modelling approach that we adopt in our fiducial mock surveys has the significant advantage of guaranteeing internal self-consistency: for a known halo $v_\mathrm{max}$ function (that realised in a given simulation), the corresponding $w_{50}$ function is unambiguously predicted and, given knowledge of the survey selection function, predictions of source counts are also unambiguous. This therefore seems like a much more robust way to test the consistency of cosmological models with observation.

This is not to say that forward-modelling the \HIWF{} from a simulation is without caveats. The prediction for the line widths is limited by how faithfully the simulation models the gas ionization state and kinematics in galaxies which depends on physical processes that are notoriously difficult to model, such as supernova feedback. It can be hoped, however, that whether a given simulation is faithful enough in its reproduction of the detailed internal structure of galaxies can be assessed on the basis of comparison with similarly detailed observations of galaxies.

Another hurdle to using this forward-modelling approach for constraining cosmology is that it is challenging to show that a prediction for the \HIWF{} is unique to a particular cosmological model and cannot be reproduced in a different cosmogony. The prediction is unlikely to be completely degenerate -- it seems unlikely that if two cosmological models resulted in identical predictions for the \HIWF{} that they would also predict otherwise indistinguishable populations of galaxies -- so again there seems to be hope for progress through folding in additional observational constraints.

\section{Conclusions}
\label{sec:conclusions}

Our analysis suggests that if dark matter is warm (with an equivalent thermal relic mass of about $10\,\mathrm{keV}$ or less) then it is likely to leave a signature in the number counts as a function of spectral line width in 21\nobreakdash-cm surveys. The (marginalised) \HIWF{} is likely to be degenerate, or nearly degenerate, for combinations of different dark matter models and assumptions about galaxy formation physics, but this degeneracy can almost certainly be broken by considering the joint distribution of the spectral line width and other observables. For example, two combinations of different dark matter models and galaxy formation models that predict the same \HIWF{} are unlikely to also predict identical galaxy stellar mass functions, and even less likely to predict the same joint distribution of stellar mass and \HI{} line width. This opens up an avenue to attempt to place quantitative constraints on WDM cosmological models in a region of parameter space competitive with other methods, such as those obtained from Ly\nobreakdash-$\alpha$ forest measurements \citep[e.g.][]{2002MNRAS.333..544H,2005PhRvD..71f3534V,2006PhRvL..97g1301V,2006PhRvL..97s1303S,2018PhRvD..98h3540M,2020JCAP...04..038P,2021MNRAS.502.2356G}.

As we have discussed above, the EAGLE galaxy formation model is rather limited in its treatment of \HI{} by, amongst other shortcomings, the pressure floor imposed on the gas in the model \citep[see][for a detailed discussion of the rationale for this, and of recent advances that obviate the need for it]{2024MNRAS.528.2930P}, the lack of local ionizing sources \citep{2020MNRAS.499.5732K,2020MNRAS.499.5862L}, and the mismatch in energetics between feedback events in nature and those modelled by the simulation \citep[][especially sec.~3.1]{2015MNRAS.450.1937C}. Our view is that this choice in modelling wreaks sufficient havoc at the low-\HI{} mass and low-$w_{50}$ end of the galaxy population as to make quantitative constraints on WDM models based on the \HIWF{} using these simulations uncertain. We are optimistic that very recent \citep[e.g.][]{2023MNRAS.522.3831F} and future simulations, such as those with the forthcoming \textsc{colibre} model (Schaye et al. in preparation), that explicitly follow gas to much lower temperatures and higher densities (while still recreating a realistic and statistically useful galaxy population) will offer opportunities for progress with the approach outlined in this work.

The \AA{} survey is the current gold standard for work on the \HIWF{}, but will very soon be superseded by new surveys including WALLABY \citep{2020Ap&SS.365..118K} on ASKAP\footnote{Australian Square Kilometer Array Pathfinder.} and CRAFTS \citep{2019SCPMA..6259506Z} on FAST\footnote{Five-hundred-metre Aperture Spherical radio Telescope.}, and eventually the Square Kilometre Array \citep[see][for some discussion around constraining dark matter models including WDM using SKA measurements of the \HIWF{}]{2015aska.confE.133P}. Notably, the WALLABY survey has somewhat better sensitivity than \AA{} and when completed will cover triple the area on the sky, with projections of up to a factor of $25$ more detections expected. CRAFTS is projected to detect up to $6\times 10^{5}$ galaxies in a $20\,000\,\mathrm{deg}^2$ field out to $z<0.35$. Of the WALLABY detections, about $20\,000$ (the total number of `Code I' sources detected in \AA{}) will be at least marginally spatially resolved. Amongst many advantages, this offers the opportunity to identify galaxies where the gas does not sample the flat part of the rotation curve -- an important systematic effect, as we have seen above.

The statistical (i.e. counting) uncertainties on the \HIWF{} as determined from \AA{} are already small \citep[see][]{2022MNRAS.509.3268O} and are expected to shrink by a factor of $5$ with WALLABY. This places us firmly in a systematic uncertainty-dominated regime; to make full use of the available data, the burden is on theoretical predictions to control for these at a level that allows for sufficient confidence in constraints on WDM models.

\section*{Acknowledgements}\label{sec:acknowledgements}

We thanks Adrian Jenkins for assistance with preparing initial conditions for the simulations used in this work, and the EAGLE and E\nobreakdash-MOSAICS collaborations for making their simulations available to us.

KAO acknowledges support by the Royal Society through Dorothy Hodgkin Fellowship DHF/R1/231105 and by STFC through grant ST/T000244/1. KAO \& CSF acknowledge support by the European Research Council (ERC) through an Advanced Investigator grant to C.S. Frenk, DMIDAS (GA 786910). JP is supported by the Australian government through the Australian Research Council's Discovery Projects funding scheme (DP220101863). This work used the DiRAC@Durham facility managed by the Institute for Computational Cosmology on behalf of the STFC DiRAC HPC Facility (www.dirac.ac.uk). The equipment was funded by BEIS capital funding via STFC capital grants ST/K00042X/1, ST/P002293/1, ST/R002371/1 and ST/S002502/1, Durham University and STFC operations grant ST/R000832/1. DiRAC is part of the National e-Infrastructure. The study also made use of the Prospero HPC facility at Liverpool John Moores University. This research has made use of NASA's Astrophysics Data System.

\section*{Software}

This work has made use of the following software packages: \textsc{numpy} \citep{2020Natur.585..357H}, \textsc{scipy} \citep{2020NatMe..17..261V}, \textsc{astropy} \citep{2013A&A...558A..33A,2022ApJ...935..167A}, \textsc{matplotlib} \citep{2007CSE.....9...90H,2023zndo...7697899C}, \textsc{martini} \citep{2019MNRAS.482..821O,2019ascl.soft11005O,2024JOSS....9.6860O} and \textsc{pandas} \citep{mckinney-proc-scipy-2010,2023zndo...7549438T}.

\section*{Data availability}

The output of the simulations used in this work are available as follows:
\begin{itemize}
\item {\EAGLECDM{}: refer to public data release \citep{2016A&C....15...72M,2017arXiv170609899T}.}
\item {\EMOSAICSCDM{}: available on reasonable request to RAC (R.A.Crain@ljmu.ac.uk).}
\item {\EAGLEWDM{}: Available on reasonable request to KAO (kyle.a.oman@durham.ac.uk).}
\item {\EMOSAICSWDM{}: available on reasonable request to RAC (R.A.Crain@ljmu.ac.uk).}
\end{itemize}

Tabulated values for galaxies in our mock surveys as described in Sec.~\ref{subsec:methods-mocks} are available from KAO on reasonable request.

\bibliography{paper}



\label{lastpage}
\end{document}